\documentstyle[prl,aps,psfig,floats,twocolumn]{revtex}
\begin{document}
\draft
\twocolumn[\hsize\textwidth\columnwidth\hsize\csname@twocolumnfalse\endcsname
\title{Scale-free Networks without Growth or Preferential Attachment: 
Good get Richer.}
\author{G. Caldarelli$^1$, A. Capocci $^2$,  
P. De Los Rios $^{3,4}$, and M. A. Mu\~noz $^5$ .}
\address{
$^1$INFM UdR ROMA1 Dip. Fisica, Universit\`a di Roma 
``La Sapienza'' P.le A. Moro 2 00185 Roma, Italy.\\
$^2$ D\'epartement de Physique, Universit\'e de Fribourg-P\'erolles,
CH-1700 Fribourg Switzerland. 
$^3$Institut de Physique Th\'eorique, Universit\'e de Lausanne
CH-1004 Lausanne Switzerland.\\ 
$^4$INFM UdR Politecnico di Torino, Corso Duca degli Abruzzi 24,
10129 Torino Italy.\\
$^5$ Instituto de F{\'\i}sica Te\'orica y Compt. Carlos I,
Universidad de Granada, Facultad de Ciencias, 18071-Granada, Spain.} 
\date{\today} 
\maketitle 
\begin{abstract} 
A new mechanism leading to scale-free networks is proposed in this letter. 
It is shown that in many cases of interest, the connectivity power-law behavior 
is neither related to dynamical properties nor to preferential attachment.
Instead, we show that without increasing the number of vertices in time 
and without applying the so called {\it ``rich-get-richer''} condition we 
obtain networks whose statistical properties are scale-free.
Assigning a quenched fitness value $x_i$ to every vertex,
and drawing links among vertices with a probability depending 
on the fitnesses of the two involved sites, 
gives rise to what we call a {\it ``good-get-richer''} 
mechanism, in which sites with larger fitness are more likely to 
become hubs (i.e., to be highly connected).
This procedure generates power-law
behaviors for various fitness distributions and attaching rules. 
\end{abstract} 
\pacs{89.20.Hh, 05.70.La, 05.65+b} 
] 
\narrowtext 

Complex networks\cite{stro,revmod,mendes} are attracting much interest
as one of the most promising tools to describe 
a large variety of social \cite{social}, biological \cite{bio,FW1}, and
technological systems, as the Internet\cite{FFF99,CMP00,Ale1,Ale2} or the World
Wide Web (WWW) \cite{AJB99,KRRT99}.
Networks are abstract mathematical objects composed by 
vertices (sites) connected by arcs (links).
In the aforementioned examples, vertices can represent people, proteins, 
species, routers or html documents, while arcs correspond to
acquaintances, physical interactions, predation relationships, 
cable connections or hyperlinks respectively.
In recent developments, Scale Free (SF) networks,
 have emerged in many different contexts, as the WWW,
the Internet,  e-mail and scientific-citation networks, protein and gene 
interaction networks, etc, and have become paradigmatic 
\cite{revmod,mendes}.
 In all these examples, the degree $k$ of a vertex, 
{\it i.e.} the number of arcs linking it to other vertices, 
is power-law distributed, $P(k) \sim k^{-\gamma}$.
SF networks also present the, so called, {\it small-world} phenomenon \cite{sw},
that is, by few selected jumps 
(that can be either short- and long range) it is possible to reach very 
different regions of the system and apparently distant environments. 

To understand how SF networks arise, 
the concepts of {\it growing networks} and
of {\it preferential attachment} have been introduced \cite{revmod}.
In particular, in the best known SF network model, i.e. the 
Barab\'asi-Albert (BA) one,  networks  grow at a constant rate 
(modeling the fact that new Web pages are continuously created, 
new proteins emerge by mutation, and so forth), and
new vertices are attached to older ones with a probability $p(k)$ 
which is a (linearly \cite{krap}) growing function of the number of 
preexisting links, $k$, at every site. 
In this way,  a highly connected vertex is more likely to receive 
further links from newly arriving vertices: 
this is the so-called "rich get richer" rule.

In some other, recently proposed, models of protein interaction networks
\cite{Amos,Sole} 
and of the WWW\cite{KRRT99}, new vertices (proteins and Web pages respectively)
are added by  copying (replicating) existing vertices, borrowing some of their 
links and adding some new others.
It has been shown that this mechanism leads also to an {\it effective} preferential 
attachment mechanism.   

Yet, although in some contexts preferential attachment can be a 
very reasonable assumption, in many others it is certainly not. 
In particular, in some situations, the information about the degree of each and 
every single vertex is not available to newly added sites, 
neither in a direct nor in an effective way.
Instead, it is reasonable that two vertices are connected when the link 
creates a mutual benefit (here we restrict ourselves to bidirectional links) 
depending on some of their intrinsic properties 
(authoritativeness, friendship, social success, 
scientific relevance, interaction strength, etc). 
Therefore, it is reasonable to expect that for some of these systems
the $P(k)$ scale free behavior (when existing)
has an origin unrelated to preferential attachment. 

In order to explore this simple idea, we propose the following
network-building algorithm:
\begin{itemize}
\item Start by creating a total (large) number $N$ of vertices.
 At every vertex $i$ a fitness
$x_i$, which is a real number measuring its importance or rank, is assigned.
Fitnesses are random numbers taken from a given probability distribution
$\rho(x)$.
  
\item For every couple of vertices, $i,j$, a link is drawn with a probability
$f(x_i,x_j)$ ($f$ a symmetric function of its arguments)
depending on the ``importance'' of both vertices, {\it i.e.} on $x_i, x_j$. 
\end{itemize}
Some remarks are in order before proceeding further:

i) The concept of ``vertex-importance'' or fitness has been  
already introduced successfully in the field of complex networks, 
but as an additional ingredient on top of the BA network \cite{bia}. 
Contrarily, here we put the emphasis on fitness itself, by
eliminating the preferential attachment rule.
ii) A trivial realization of the above rules is the standard 
Erd\H{o}s R\'enyi model \cite{ER61}, 
where $f(x_i,x_j)$ is constant and equal to $p$ for all vertex couples.
This particular choice does not produce SF networks, but in what follows we will show
that other realizations of the general rules do so. 
iii) The model, as defined, is static, but it can be 
straightforwardly be considered a
dynamic one
 by adding new vertices at every time step and linking them to the existing ones
according to the above attaching rule.  
iv) It is also easy to generalize the model to include asymmetric or directed links.
v) A somehow similar static model was studied by Goh {\it et. al.} \cite{Goh}. 

A general expression for $P(k)$ can be easily derived. 
Indeed, the mean degree of a vertex of fitness $x$ is simply
\begin{equation}
k(x) = N \int_0^\infty f(x,y) \rho(y) dy = N F(x)
\label{general formula}
\end{equation}
(with  $x_i \in (0,\infty)$).
Assuming $F(x)$ to be a monotonous function of $x$, and for large enough N,
we have the simple relation
\begin{equation}
P(k) = \rho \left[ F^{-1}\left(\frac{k}{N}\right)\right]
\frac{d}{dk}F^{-1}\left(\frac{k}{N}\right).
\label{inversion}
\end{equation}
 For finite values of $N$ corrections to this equation emerge \cite{Redner}. 
As a particular example, consider $f(x_i,x_j)=(x_i x_j)/x_M^2$ 
where $x_M$ is the largest value of $x$ in the network. Then 
\begin{equation}
k(x) = \frac{N x}{x_M^2} \int_0^\infty y \rho(y) dy = N \frac{<x> x}{x_M^2}
\label{degree vs x}
\end{equation}
and we have the simple relation 
\begin{equation}
P(k) = \frac{x_M^2}{N <x>} \rho\left(\frac{x_M^2}{N <x>} k\right).
\label{P(k)}
\end{equation}
 A particularly simple realization of the model emerges if we consider
power-law distributed fitnesses. This choice can be naturally justified 
by arguing that power-laws appear rather generically in many contexts
 when one ranks, for example, people according to their incomes or
cities according to their population, etc. 
This is the so-called Zipf law which
establishes that the rank $R(x)$ behaves as  $R(x) \propto x^{-\alpha} $ in a
quite universal fashion \cite{Zipf}.
The reason for the ubiquitous presence of the Zipf law yields on the
multiplicative nature of the intrinsic fluctuations which generically leads to flat 
distributions in logarithmic space and, consequently, to power-laws \cite{Zipf}.

\begin{figure}
\centerline{\psfig{file=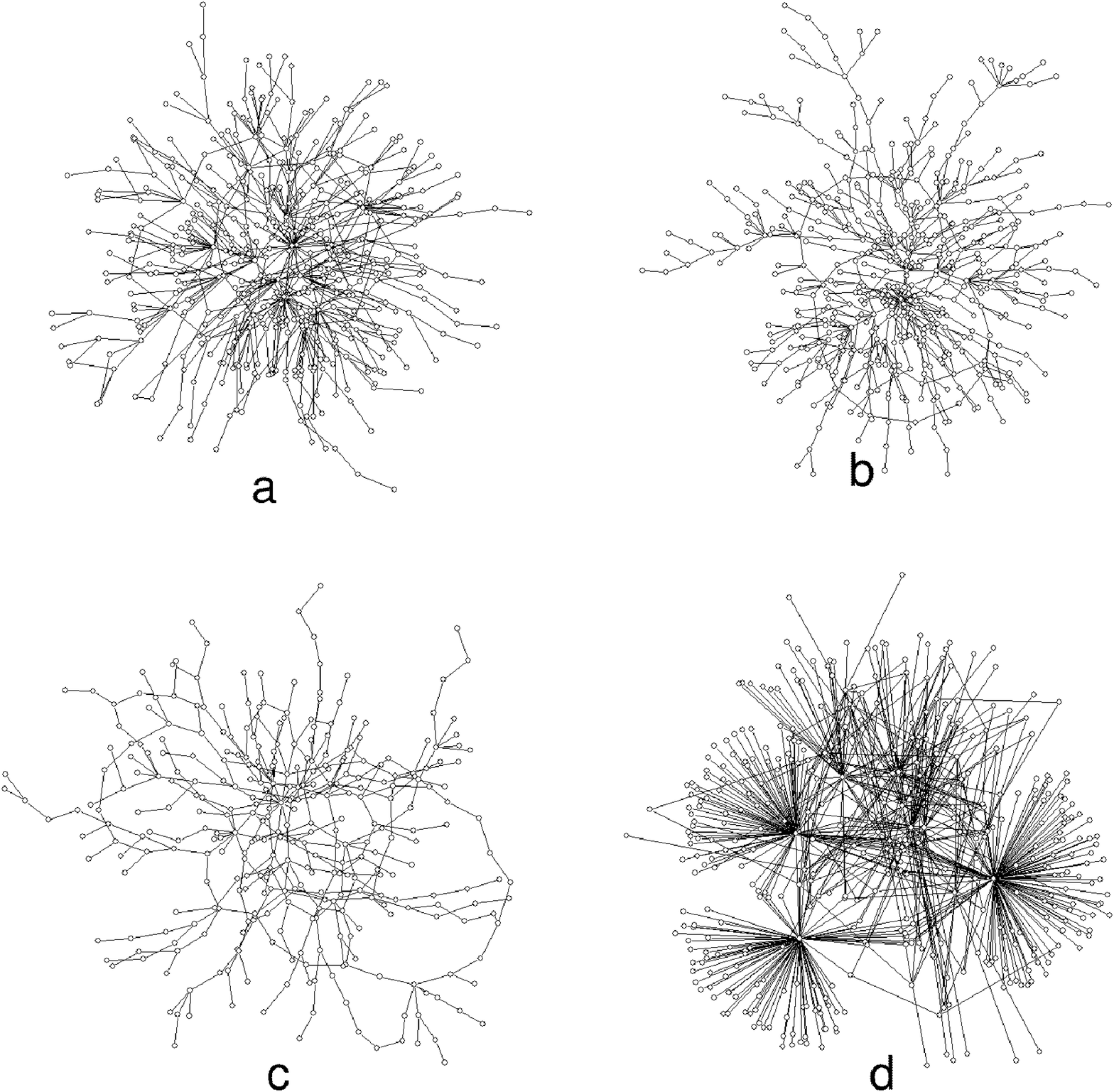,width=8cm}}
\caption{Graph representations of four networks produced respectively with
({\bf a}) $\beta=2.5$, ({\bf b}) $\beta=3.0$, ({\bf c}) $\beta=4.0$, and
({\bf d}) $\rho(x)=e^{-x}$ with a threshold rule. Graphs have been produced
with the Pajek software.}
\label{Fig1}
\end{figure}

\begin{figure} 
\centerline{\psfig{file=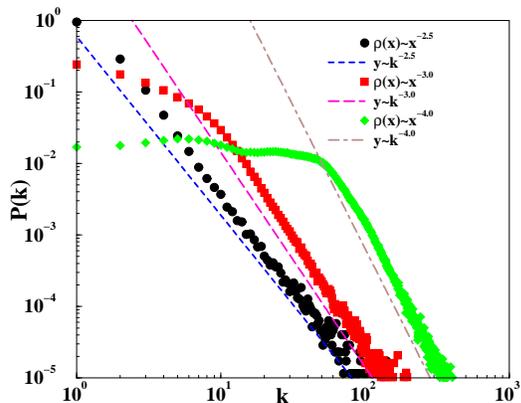,width=7cm}}
\caption{Degree distribution for networks generated using 
$\rho(x) \sim x^{-\beta}$
with $\beta=2.5,3,4$, and
$f(x_i,x_j)=(x_i x_j)/x_M^2$;
 power-laws with the corresponding analytical values 
are explicitely drawn in straight lines. 
Data obtained for networks with $10^4$ vertices, 
averaged over $100$ realizations.}
\label{Fig2}
\end{figure}

Clearly, if $\rho(x) \sim x^{-\beta}$  (Zipf's behavior, with $\beta = 1+1/\alpha$ 
\cite{Zipf}) then, using eq.(\ref{P(k)}), also the degree distribution $P(k)$ 
is a power-law and the network shows SF behavior.   
In Fig.\ref{Fig1} we show the degree distributions from
simulations with $\beta=2.5,3,4$ (corresponding to
Zipf exponents $\alpha=2/3,1/2,1/3$); the asymptotic behavior
is, in all cases, well described by eq.(\ref{P(k)}).
This result is hardly surprising: 
from SF fitnesses we generate SF networks, but still it provides a new generic
path to SF networks and takes into account the widespread occurrence of the Zipf's
behavior in nature.
In order to extend this result and check whether SF networks can be generated even 
when $\rho(x)$ is not SF itself,
we consider an exponential distribution of fitnesses,
 $\rho(x) = e^{-x}$ (representing a random, Poisson distribution)
 and  $f(x_i,x_j) = \theta(x_i+x_j-z)$, where $\theta(x)$ is 
the usual Heaviside step function.
 This represents processes where two vertices are 
linked only if the sum of their fitnesses is larger than a given {\it threshold} $z(N)$.
Using these rules we obtain analytically (and confirm in
computer simulations) that $P(k) \sim k^{-2}$ \cite{footnote}.
This leads to the non-trivial result that {\it even non scale-free fitness 
distributions can generate scale-free networks} (see Fig.\ref{Fig2}).
Also different implementations of the threshold rule, such as
$f(x_i,x_j) = \theta(x_i^n+x_j^n-z^n)$ (where $n$ is an integer number)
give rise to the same inverse square behavior 
(although, in some cases, with logarithmic corrections).

In a future publication we will explore, in a more systematic way,
the necessary and sufficient conditions for the fitness distribution
and attaching rule under which well-behaved SF networks are generated.

Let us stress that the model, as defined, has a diverging average connectivity 
in the large N limit, as can be easily inferred from Eq.(\ref{general formula}); 
{\it i.e.} it is severely {\it accelerated} \cite{acc}.
Nevertheless we can introduce in a rather natural way
an upper cut-off accounting for the fact that every site has a limited information 
on the rest of the world and, therefore, connection is attempted with a finite number,
$m$, of different sites. 
Alternatively, vertices can be linked with the above rule 
and, after that, links are kept with probability $p$ (so that, for example, $pN=m$).
By including this modification, the $N$ factor in Eq.(\ref{general formula}),
is substituted by $m$, and the connectivity is finite in the thermodynamic limit.
In order to generate different  accelerated networks
(with the averaged connectivity not reaching a stationary value 
but growing with $N$ in different possible ways \cite{acc}) other
selection rules can be easily implemented.

To have a more extensive picture of the nature of the networks under
consideration, we have studied the following topological properties 
\cite{mendes},
interest in which has been triggered by  recent
studies on the Internet structure \cite{Ale2,Ale3}:
\begin{itemize}
\item The average distance $ < d >$, measuring the average
minimum number of arcs needed to connect two given sites.
\item 
The {\it average neighbor connectivity}  $k_{nn}(k)$, measuring
the average degree of vertices neighbor of a k-degree vertex. 
\item
The {\it clustering coefficient} $c(k)$ 
that measures the degree of interconnectivity of
nearest neighbors of k-degree vertices. 
More specifically the clustering coefficient $c_i$ of a vertex $i$,
whose degree is $k_i$, is the ratio between the number of 
edges $e_i$ in the subgraph identified by its neighbors and the maximum 
possible number of edges in the subgraph. That is $c_i = 2e_i/k_i(k_i-1)$
\cite{revmod}.
 $c(k)$ is obtained by averaging  $c_i$ for all vertices with fixed degree $k$.
\item 
The probability distribution of the 
{\it betweenness}, $b_i$, defined as the total number of minimum 
paths between any couple of vertices in the network passing 
through vertex $i$~\cite{MEJ}. 
This quantity gives a measure of the amount of traffic passing through a vertex. 
We studied, as in the aforementioned papers, both the probability distribution 
$P(b)$ and its first moment  $<b>/N$.
\end{itemize}

Computer simulations of our model show that networks with 
power-law distributed fitnesses, 
and different values of  $\beta$, 
show nearly constant $k_{nn}(k)$'s and $c(k)$'s,
just as occurs for the original BA model   \cite{revmod}. 
The distribution of betweenness decays as a power law 
with an exponent  $\gamma_b \approx 2.2$ for 
$\gamma=2.5$ and $\gamma=3$, and $\gamma_b \approx 2.6$ 
for $\gamma=4$.
This is in good agreement with what conjectured in Ref. \cite{Goh}:
all networks with $3 \geq \gamma >2$ 
can be classified in only two groups according to the value of 
$\gamma_b$ ( $\gamma_b=2$ and $\gamma_b=2.2$, respectively), while 
for larger values of $\gamma$, larger non-universal values 
of $\gamma_b$ are reported. 

The exponential case behaves in a different way: 
for a network of size $N=10^4$, $z=10$, and $m=N$ we find
 $<d> = 2$,  $<c> \simeq 0.1$ and $<b>/N\simeq0.1$, but
a power-law  behavior is found for the clustering magnitudes, 
{\it i.e.} $<k_{nn}>\propto k^{-0.85}$ and $c(k)\propto k^{-1.6}$. 
The betweenness distribution instead, shows an unexpected behavior, 
giving a power-law tail with an exponent 
$\gamma_b \approx 1.45$ (see Fig.4). 
It is worth remarking that our model having  $\gamma=2$ is not
included in the  previously discussed classification of betweenness
exponents \cite{Goh}.

Having explored the most basic properties of the model and some particular 
realizations, let us comment now on possible applications.

E-mail networks \cite{ebel} are a good candidate to be 
represented by our model.
In this case growth may occur, but agents 
(e-mail senders) do not have any access or 
knowledge of the degree of the receivers. Rather than preferential
attachment there should be some intrinsic feature of the 
receiver playing a role in the phenomenon.

To further emphasize the utility of this new mechanism let us
mention the following possibility: one can imagine situations where
a Poisson network is seen as SF just because the exploration method implicitly
implements a probabilistic rule depending on the fitnesses 
(this applies for example when links are detected by "picking" them one by 
one, but not if the network is explored by crawling on it). 
Let us think, for example, of the case with threshold type of attaching rule. 
If only links with corresponding fitnesses over threshold are ``seen'' by
the exploration method then, for example, an  Erd\H{o}s R\'enyi network 
with exponentially distributed fitnesses can be seen as SF (with, obviously,
a connectivity upper cutt-off related to the maximum connectivity of the
underlying network; in cases in which  this connectivity is  high, one can
generate hubs in the ``apparent'' SF network).
In particular this scenario could be of relevance to protein networks. 
Let us argue why.

The way comprehensive protein networks have been obtained to
date is through a bait-prey method, named "two-hybrid" method: 
two proteins are hybridized with two fragments of a transcription 
factor (a protein that binding to a gene
promotes its transcription into the corresponding RNA).
The spliced promoter does not bind to the gene, transcription is 
inhibited and the corresponding RNA is absent.
Yet, if the two proteins interact they bring together the two promoter fragments
allowing it to bind to the gene and transcription to start.
The presence of the corresponding RNA signals the interaction between the
two proteins.
We can imagine that the interaction strength between
the two proteins has to be above
a given threshold, else the typical promoter binding time will be too short
for the RNA polymerase to bind to the gene
and initiate transcription. In turn it is reasonable to assume that the
interaction strength is a function of some properties of the two proteins 
(such as, for example, their hydrophobicity, or their Accessible 
Surface Area). This possibility has still to be checked through an 
analysis of the detailed physics behind the two-hybrid method.

In summary, we have presented an alternative model
to justify the ubiquity of SF networks in nature. 
It is a natural generalization of the standard Erd\H{o}s-R\'enyi.
The main result is that emergence of SF properties is not 
necessarily linked to the ingredients of growth and preferential attachment. 
Instead, static structures characterized by quenched disorder 
(for different disorder distributions) 
and threshold phenomena, may generate effects very similar to those
measured in the real data. In particular we recover the power-law behavior of 
degree, betweenness, and clustering-coefficient distributions. 
We believe that this model is particularly suitable for 
situations where the degree value of nodes is not publicly available.

We thank R. Pastor-Satorras for very useful comments and suggestions,
and P. Hurtado for a critical reading of the manuscript.
The authors wish to thank  FET Open Project IST-2001-33555 COSIN, 
P.D.L.R. thanks the OFES-Bern (grant No. 02.0234), 
and M.A.M. acknowledges financial support
from the MCYT (grant No. BFM2001-2841). 

\begin{figure}
\centerline{\psfig{file=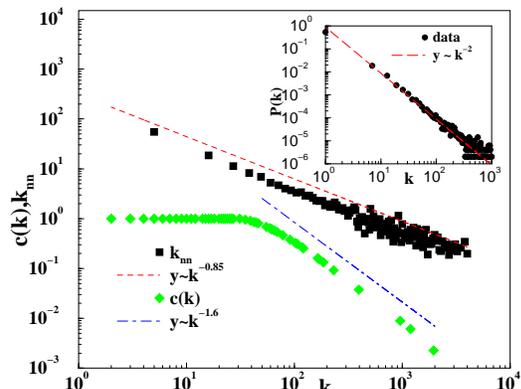,width=7cm}}
\caption{
Average neighbor connectivity $k_{nn}$ against $k$,
for networks generated using $\rho(x) \sim e^{-x}$
and a threshold rule. Results are
averaged over $100$ realizations of size $10^4$.
Inset: the degree distribution $P(k)$;
the straight line corresponds to $k^{-2}$.}
\label{Fig3}
\end{figure}

\begin{figure}
\centerline{\psfig{file=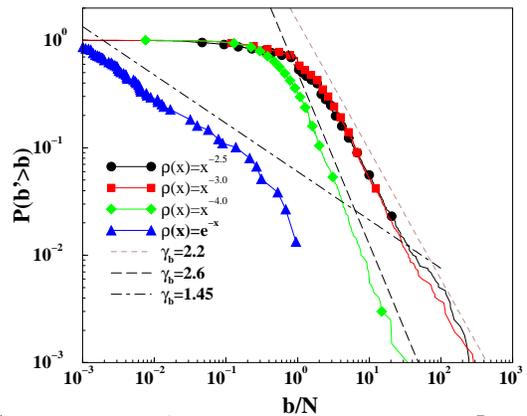,width=7cm}}
\caption{Integrated betweenness distribution,
$P(b'>b) \propto b^{1-\gamma_b}$,
 for different fitness distributions.}
\label{Fig4}
\end{figure}


\begin{thebibliography}{99}
\bibitem{footnote} 
Being more precise, we obtain a power law plus a delta function
of amplitude $exp(-z)$ at
$K=N$, corresponding to the fact that sites with $x_i$ above threshold
are connected to every other site. 
This is irrelevant for large enough z's and N's.
\end{thebibliography}
\end{document}